\documentclass[manuscript]{aastex}

\shorttitle{Fast Rotation of Near-Earth Object 2011~$\mathrm{XA_3}$}
\shortauthors{Urakawai et al.}

\begin{document}


\title{Fast Rotation of a Sub-km-sized Near-Earth Object 2011~$\mathrm{XA_3}$}


\author{SEITARO URAKAWA}
\affil{Bisei Spaceguard Center, Japan Spaceguard Association, 1716-3 Okura, Bisei, Ibara, Okayama 714-1411, Japan}
\email{urakawa@spaceguard.or.jp}

\author{KATSUHITO OHTSUKA}
\affil{Tokyo Meteor Network, 1-27-5, Tokyo 155-0032, Japan}

\author{SHINSUKE ABE}
\affil{Department of Aerospace Engineering, College of Science and Technology, Nihon University, 7-24-1, Narashinodai, Funabashi, Chiba 274-8501, Japan}

\author{TAKASHI ITO}
\affil{National Astronomical Observatory of Japan, 2-21-1 Osawa, Mitaka, Tokyo 181-8588, Japan}

\and
\author{TOMOKI NAKAMURA}
\affil{Department of Earth and Planetary Material Sciences, Graduate School of Science, Tohoku University, Aoba, Sendai, Miyagi 980-8578, Japan}



\received {2013/12/13}
\accepted{2014/3/3}
\begin{abstract}
We present lightcurve observations and their multiband photometry for near-Earth object (NEO) 2011~$\mathrm{XA_3}$. The lightcurve has shown a periodicity of 0.0304 $\pm$ 0.0003 day (= 43.8 $\pm$ 0.4 min). The fast rotation shows that 2011~$\mathrm{XA_3}$ is in a state of tension  (i.e. a monolithic asteroid) and cannot be held together by self-gravitation. Moreover, the multiband photometric analysis indicates that the taxonomic class of 2011~$\mathrm{XA_3}$ is S-complex, or V-type. Its estimated effective diameter is 225 $\pm$ 97~m (S-complex) and 166 $\pm$ 63~m (V-type), respectively. Therefore, 2011~$\mathrm{XA_3}$ is a candidate for the second-largest fast-rotating monolithic asteroid. Moreover, the orbital parameters of 2011~$\mathrm{XA_3}$ are apparently similar to those of NEO (3200) Phaethon but F/B-type. We computed the orbital evolutions of  2011~$\mathrm{XA_3}$ and Phaethon. However, the results of the computation and distinct taxonomy indicate that neither of the asteroids is of common origin. 
\end{abstract}





\section{Introduction}
The physical properties of asteroids provide us with important clues to clarify the compositions, strengths, and impact history of planetesimals that were formed in the early solar system. One representative methods for investigating the physical properties of asteroids is their lightcurve observations. Lightcurve observations are able to deduce the rotational status and shapes of asteroids. Past observations found that diameters of most asteroids rotating shorter than 2.2~h were smaller than 200~m \citep{Pravec00}. The fast-rotating asteroids have structurally significant tensile strength (so-called monolithic asteroids) because such asteroids need to overcome their own centrifugal force. On the other hand, the slow rotating asteroids larger than 200~m in diameter maintain themselves by their gravity or the cohesive force of bonded aggregates \citep{Ric09}. Some such asteroids are structurally rubble pile \citep{Abe06}. 
  The near-Earth object (NEO), 2001~OE$_{84}$, has the largest diameter among fast-rotating asteroids whose rotational periods are reasonably determined in the Asteroid Lightcurve Database (LCDB) with quality code U$\ge$2 \citep{Warner09}. When the geometric albedo $p_{V}$ is 0.18, its diameter and rotational period are $\sim 700$~m and 29.1909~min, respectively \citep{Pravec02}. \cite{Hicks09} has reported that the diameter and rotational period of the NEO 2001~FE$_{90}$ are 200~m and 28.66~min, assuming the geometric albedo $p_{V}$ of 0.25. The diameter and rotational period of 2001~VF$_2$ are 1.39~h and 145--665~m \citep{Her11}. Despite searches having been devoted to their detection, few fast-rotating asteroids clearly larger than 200 m have been found \citep{Pravec00b, Whiteley02, Kwiat10a, Kwiat10b}. A numerical simulation indicates that fast-rotating asteroids are provided by the collisional disruption of parent bodies \citep{Asphaug99}. The relationship between the diameter and rotational period of ejecta in a laboratory impact experiment is similar to that of monolithic asteroids \citep{Kado09}. Moreover, explorations by NEAR and Hayabusa spacecrafts have discovered that the shapes of boulders on NEOs (433) Eros (mean diameter of 16.84~km) and (25143) Itokawa (mean diameter of 0.33~km) resembled those of monolithic asteroids \citep{Michi10}. Therefore, monolithic asteroids are thought to be generated by impact craterings or catastrophic disruptions on the parent bodies. 
  The Rosetta and the NEAR spacecraft have determined the size frequency distributions of boulders on the main belt asteroid (MBA) (21) Lutetia (mean diameter of 95.76~km) and Eros \citep{Thomas01, Kup12}. In the case of Lutetia, the largest boulder is $\sim 300$~m in diameter. The power law index $\alpha$ of the cumulative size-frequency distribution, $N(>D)$ $\propto$ $D$$^{-\alpha}$, is about -5 and becomes shallower in regions smaller than 150~m, where $N$ is the number of boulders and $D$ is the diameter of boulders. A similar trend in slope is also shown for Eros.  When we hypothesize that the size frequency distributions of the boulders on Lutetia and Eros are recognized as the typical size frequency distribution of impact ejecta,  some ejecta smaller than 150~m can escape from the parent object that are likely to become small and fast-rotating asteroids. To confirm such a hypothesis, we need to show clearly how the fast-rotating asteroids inhabit from 150 m or less to subkilometer size by examining gradual but steady observational data accumulations. \\
  
           The purpose of our study is to obtain the rotational period, the taxonomic class for NEO 2011~$\mathrm{XA_3}$. Since apparent brightness increases when NEOs closely encounter the Earth, the observations of NEOs are suitable for the elucidation of physical properties of subkilometer-sized asteroids. Revealing the taxonomic class confines the albedo, and increases the estimate accuracy for the diameter. 2011~$\mathrm{XA_3}$ was discovered by Pan-STARRS (the Panoramic Survey Telescope \& Rapid Response System) on December 15,  2011. The absolute magnitude $H$ = 20.402 $\pm$ 0.399 (JPL Small-Body database\footnote{http://ssd.jpl.nasa.gov/sbdb.cgi}) indicates that the object is a subkilometer-sized asteroid. Based on more than 100 astrometric observations for this asteroid, the orbital parameters are reduced to $a$ = 1.48 AU, $e$ = 0.93, $i$ = $28\fdg1$, $\Omega$ = $273\fdg6$ and $\omega$ = $323\fdg8$. Its orbit is apparently similar to those of NEO (3200) Phaethon, $a$ = 1.27 AU, $e$ = 0.89, $i$ = $22\fdg2$, $\Omega$ = $265\fdg3$ and $\omega$ = $322\fdg1$, though the semi-major axis has a difference. Phaethon is the parent body of the Geminids, one of the most intense meteor showers throughout the year. Moreover, the Phaethon Geminid Complex (hearafter, PGC) like 2005 UD and 1999 YC,  which are objects dynamically linked with Phaethon and the Geminids, are proposed \citep{Ohtsuka06, Ohtsuka08}. In fact, B/F/C taxonomy of 2005 UD and 1999 YC is shared by Phaethon \citep{Je06,Kinoshita07, Ka08}. We also investigate whether 2011~$\mathrm{XA_3}$ is a member of PGC by the multiband photometry and by calculating the orbital motion of both 2011~$\mathrm{XA_3}$ and Phaethon. In this paper, we deal with the following: In Section 2, we describe the observations and their data reduction. In Section 3, we mention the results of rotational period, taxonomic class, and the orbital simulation.  In Section 4, we discuss the relationship between 2011~$\mathrm{XA_3}$ and Phaethon. Moreover, we focus on the heating effect due to the close perihelion distance of 2011~$\mathrm{XA_3}$. Finally, we summarize the physical properties of 2011~$\mathrm{XA_3}$ and mention the observation efficiency of subkilometer-sized NEO using small- and medium-aperture telescopes.

\section{Observations and data reductions}
\subsection{Observations}
We conducted the photometric observation of 2011~$\mathrm{XA_3}$ using the 1.0~m f/3 telescope and the 0.5~m f/2 telescope at the Bisei Spaceguard Center (BSGC)\footnote{BSGC is administrated by the Japan Space Forum.}. The observational circumstances and the states of 2011~$\mathrm{XA_3}$ are listed in Table 1 and 2, respectively. Both telescopes were operated with the non-sidereal tracking mode. The detector of 1.0~m telescope consisted of four CCD chips with 4096 $\times$ 2048 pixels. We used one CCD chip to obtain as many images as possible by shortening the processing time. The field of view (FOV) for one CCD chip is $1\fdg14$ $\times$ $0\fdg57$ with a pixel resolution of  $1\farcs0$. The observations on December 16, 2011 focused on the astrometry using the 1.0~m telescope  with the exposure time of 150 s in 2 $\times$ 2 binning. Individual images were taken with a commercially available short-pass (long-wavecut) filter indicating $W$ in Table 1 whose effective wavelength ranged from 490~nm to 910~nm. Though the observation was short term of about 45 min, we also used the data for the period analysis of the lightcurve observations. The lightcurve observations were mainly carried out on December 19, 2011 using the 0.5~m telescope. The detector of the 0.5~m telescope is Apogee U42 CCD with 2048 $\times$ 2048 pixels. The FOV is $1\fdg67$  $\times$ $1\fdg67$ with a pixel resolution of $2\farcs9$. The images were obtained with the exposure time of 120~s using $W_i$ filter \citep{Oku11}. At the same time, multiband photometry was conducted using the 1.0~m telescope with Sloan Digital Sky Survey (SDSS) $g'$, $r'$, $i'$ and $z'$ filters.  We also measured the flux of 83 standard stars from SDSS data Release 8 \citep{Aihara11}, whose stars were imaged around the same airmass with 2011~$\mathrm{XA_3}$. These objects have the $r'$-band magnitudes of about 14 mag to 16 mag and classification code 1 (= primary), quality flag 3 (= good), and object class 6 (= star). One set of observations was made using three consecutive images for each filter. The filters were changed in the following sequence: three $g'$ images (2011~$\mathrm{XA_3}$) ${\rightarrow}$ three $g'$ images (standard stars) ${\rightarrow}$ three $r'$ images (2011~$\mathrm{XA_3}$) ${\rightarrow}$ three $r'$ images (standard stars) ${\rightarrow}$ three $i'$ images (2011~$\mathrm{XA_3}$) ${\rightarrow}$ three $i'$ images (standard stars) ${\rightarrow}$ three $z'$ images (2011~$\mathrm{XA_3}$) ${\rightarrow}$ three $z'$ images (standard stars). We repeated this sequence five times. Part of the data could not be obtain due to unfavorable observing conditions between 17 h and 19 h in UT on December 19. All images were obtained with exposure time of 120~s for each filter in no binning mode.

 \subsection{Data reduction}
All images were debiased and flat-fielded. All observation time data were corrected using a light-traveled time from 2011~$\mathrm{XA_3}$ to the Earth. To obtain the lightcurve of 2011~$\mathrm{XA_3}$, we measured the raw magnitude of 2011~$\mathrm{XA_3}$ and four reference stars that were imaged simultaneously on the same field using the IRAF/APPHOT\footnote{IRAF is distributed by the National Optical Astronomy Observatory, which is operated by the Association of Universities for Research in Astronomy (AURA) under a cooperative agreement with the National Science Foundation.} package. We set the radius of aperture photometry to $\sim 1.7$ $\times$ FWHM for both 2011~$\mathrm{XA_3}$ and reference stars images, respectively. Since the reference star images are slightly elongated due to the non-sidereal tracking, the aperture radius  is larger than that of 2011~$\mathrm{XA_3}$.  We calibrated the magnitude fluctuations due to the change of atmospheric conditions using the procedures of \cite{Urakawa11}. Next, we performed the data reduction for multiband photometry. We evaluated the atmospheric extinction coefficients and conversion factors to standardize the SDSS system for each filter, in which each atmospheric extinction coefficient was calculated by the magnitude variations of the standard stars for the change in airmass. Extra-atmospheric instrumental magnitudes of both 2011~$\mathrm{XA_3}$ and the standard stars were derived using the obtained atmospheric extinction coefficient. The conversion factors were estimated by comparing the extra-atmospheric instrumental magnitudes with the cataloged magnitudes of standard stars. The brightness of 2011~$\mathrm{XA_3}$ in rotation inevitably changes during switching of the filter. We defined the time of recording the first $g' $ image as the standard time, and then we calibrated an amount of brightness change for the standard time, which was estimated by the fitting curve of the lightcurve.

\section{Results}
\subsection{Rotational period and Taxonomy}
Assuming a double-peaked lightcurve, we carry out a periodicity analysis based on the Lomb--Scargle periodgram (\citealp{Lomb76}; \citealp{Scargle82}). The power spectrum from the periodgram shows a period of 0.0304 $\pm$ 0.0003 day (= 43.8 $\pm$ 0.4 min). The folded lightcurve is shown in Figure 1. Though the total observation duration was $\sim 465$ min over two nights, we were able to detect a number of same-shaped lightcurves because of a short rotational period. Therefore, the period is sufficiently reliable to study. We also obtain the maximum amplitude of 0.68 mag by the fourth-order Fourier series fitting curve. The lightcurve has a two-humped shape at the phase of $\sim 0.7$. The two-humped shape could result from the obscuration of surface regions by local topography, because the solar phase angle was $\sim 40$$^{\circ}$ during our observation. The taxonomic class and rotational color variation for 2011~$\mathrm{XA_3}$ are investigated by two color--color diagrams (Figure 2). In order to determine the taxonomic class of 2011~$\mathrm{XA_3}$, we should avoid the surface color variability due to surface heterogeneity of the rotating asteroid. However, the fast rotation of 2011~$\mathrm{XA_3}$ makes us difficult to obtain the roughly simultaneous four-color ($g'$, $r'$, $i'$, and $z'$) data. We found all the color-indexes only between the phase 0.6 and 0.8 in the lightcurve. The graph legend of ``Phase 0.6-0.8" in Figure 2 indicates the color index of 2011~$\mathrm{XA_3}$ between the phase 0.6 and 0.8, and the graph legend of  ``Other"  indicates the color-index, which is calculated from the averaged $g'$,  $r'$, $i'$, and $z'$ magnitude except between the phases 0.6 and 0.8, even if four-color data were not obtained in the same phase. In other words, the graph legend of  ``Other" shows the color-indexes obtained by assuming that there is no surface color heterogeneity. Though the color-indexes in Figure 2 show the taxonomy of 2011~$\mathrm{XA_3}$ is plausibly to be V-type, the photometric precision is insufficient to deny the possibility of S-type. We note that the classification of taxonomy, such as Q, R, or, O-type, is difficult using this multiband photometry. Therefore we conclude that 2011~$\mathrm{XA_3}$ is V-type or S-complex (S-, Q-, R-, O-, etc.).  Since there is no clear difference between the color-index of ``Phase 0.6-0.8" and ``Other" in Figure 2, the surface color of 2011~$\mathrm{XA_3}$ is macroscopically homogeneous.  

\subsection{Diameter and Construction}
We estimated the absolute magnitude $H_{V}$ and the effective diameter for 2011~$\mathrm{XA_3}$. We deduced the apparent $r'$ magnitude of 2011~$\mathrm{XA_3}$ on 19 December, 2011 was 16.35 $\pm$ 0.05 mag at the phase where the relative magnitude was zero in Figure 1. The apparent $V$ magnitude is described in the following form \citep{Fukugita96},
 \begin{equation}
 V=r'-0.11+0.49\left(\frac{(g'-r')+0.23}{1.05}\right).
 \end{equation}
Here, for our photometric precision requirements, the difference between AB magnitude and Vega magnitude in the V-band is negligible. The reduced magnitude at the phase angle $\alpha$ is expressed as $H(\alpha) = V-5\log_{10}(R\Delta)$, where $R$ and $\Delta$ are the heliocentric and geocentric distance in AU. The absolute magnitude is expressed as a so-called $H$-$G$ function \citep{Bow89},
\begin{equation}
H_{V}=H(\alpha)+2.5\log_{10}[(1-G)\Phi_{1}(\alpha)+G\Phi_{2}(\alpha)],
\end{equation}
where $G$ is the slope parameter depending on the asteroid's taxonomy. When we apply $G$ = 0.24 $\pm$ 0.11 for  S-complex and $G$ = 0.43 $\pm$ 0.08 for V-type \citep{Warner09}, $H_{V}$ becomes 20.56 $\pm$ 0.43 mag (S-complex) and 20.84 $\pm$ 0.25 mag (V-type), respectively. An effective diameter of asteroids $D$ (in kilometer) is described as 
\begin{equation}
D = 1329\times10^{-H_{V}/5}p_{V}^{-1/2},
\end{equation}
where $p_{V}$ is geometric albedo. Assuming the albedo of 0.209 $\pm$ 0.008 for S-complex \citep{Pravec12} and 0.297 $\pm$ 0.131 for V-type \citep{Usui13}, we found the diameter of 225 $\pm$ 97~m (S-complex) and 166 $\pm$ 63~m (V-type), respectively. 

Next, we estimate the axial ratio of 2011~$\mathrm{XA_3}$. There is a relation between the lightcurve amplitude and the phase angle as follows:
 \begin{equation}
A(0) = \frac{A(\alpha)}{1+m\alpha}.
\end{equation}
Here, $A(\alpha)$ is the lightcurve amplitude at $\alpha^{\circ}$ phase angle and $m$ is the slope coefficients. In the case of an S-type asteroid, the $m$ value is 0.03 \citep{Zap90}. Moreover, the $m$ value depends on the surface roughness \citep{Ohba03}.  The $m$ value is $\sim$ 0.02 when we adopt the surface roughness 18$^{\circ}$ of Vesta as the representative for V-type \citep{Li13}. Assuming the axial ratio of 2011~$\mathrm{XA_3}$ is projected on the plane of the sky, the lightcurve amplitude is described through the lower limit to the true axial ratio of the body as 
\begin{equation}
A(0) = 2.5\log_{10}\left(\frac{a}{b}\right),
\end{equation}
where $a$ and $b$ are respectively normalized long and short axis length. The amplitude of 2011~$\mathrm{XA_3}$ is 0.68 mag and the lightcurve data are obtained at the phase angle of $\sim$ 40$^{\circ}$. Therefore, the ratio of the long axis length to the short axis length is larger than 1.3 (S-complex) and 1.4 (V-type). The obtained physical properties of 2011~$\mathrm{XA_3}$ are summarized in Table 3. The construction of fast-rotating asteroids is thought to be a monolith. We confirm the validity for 2011~$\mathrm{XA_3}$. If 2011~$\mathrm{XA_3}$ is preserved itself only by the self-gravity, the critical bulk density $\rho$ (in g~cm$^{-3}$) for a rubble pile asteroid can be written as
\begin{equation}
\rho = \left(\frac{3.3}{P}\right)^{2}(1+A(0)),
\end{equation}
where $P$ is the rotational period in hours \citep{Pravec00}. Substituting the rotational period and the lightcurve amplitude of 2011~$\mathrm{XA_3}$ into the equation (6), the bulk density $\rho$ becomes 26.3~g~cm$^{-3}$. In addition to it, a more precise calculation using a theory for cohesionless elastic-plastic solid bodies \citep{Hol04} gives a lower limit on the bulk density of 20.5~g~cm$^{-3}$ (P. Pravec, personal communication). These are an incredible values as material of asteroids. Therefore, 2011~$\mathrm{XA_3}$ is not a rubble pile but evidently a monolithic asteroid. \\

We show the rotational period and the effective diameter on Figure 3 for registered asteroids in the LCDB and 2011~$\mathrm{XA_3}$. The effective diameter of 2011~$\mathrm{XA_3}$ on Figure 3 corresponds to the mean value between the diameter of the assumed S-complex asteroids and the diameter of the assumed V-type asteroids. The range of errors designates the upper limit diameter of the assumed S-complex asteroids and the lower limit diameter of the assumed V-type asteroids. Moreover, the rotational periods and effective diameters of 2001~OE$_{84}$, 2001~FE$_{90}$, 2001~VF$_2$ and NEO 2004~VD$_{17}$ that have the possibility of being more than 200~m in diameter are plotted separately from the other asteroids in the LCDB. The nominal albedo in the LCDB for NEOs is 0.2. The absolute magnitude is deduced by assuming $G$ of 0.15. The typical uncertainty of the absolute magnitude in the LCDB is around 0.4~mag \citep{Her11}. We estimated that the diameter for 2001~OE$_{84}$ in Figure 3 based on the nominal value of the LCDB. For comparison, when we adopt the nominal albedo and $G$ for 2011~$\mathrm{XA_3}$, the effective diameter becomes $\sim$ 250~m. Indeed, the diameter includes the significant uncertainty due to the adopted albedo and absolute magnitude. We added the error bar for 2001~$\mathrm{VF_2}$ and 2001~$\mathrm{FE_{90}}$ that are monolithic asteroids and have roughly the same diameter as 2011~$\mathrm{XA_3}$. The diameter and the error range for 2001~$\mathrm{VF_2}$ are adopted from the values of \cite{Her11}.  The diameter and the error range for 2001~$\mathrm{FE_{90}}$ are estimated from $H$ = 20.1 and the A-type asteroids' albedo, $p_{V}$ = 0.282 $\pm$ 0.101 \citep{Usui13}. The rotational period and diameter for 2004 VD $_{17}$ are well determined as $\sim$ 320~m and 1.99~h by the visible-near infrared spectroscopy and the polarimetry. Though the rotational period indicates that the construction of 2004 VD$_{17}$ is categorized as a monolith, the slightly short rotation for the spin limit of 2.2~h is not deny to be partially fractured \citep{Ric02, Luise07}. Therefore, 2011~$\mathrm{XA_3}$ is a candidate of the second-largest fast-rotating monolithic asteroid behind 2001~OE$_{84}$. 

\subsection{Orbital evolutions}
We computed the orbital motions of 2011~$\mathrm{XA_3}$ and Phaethon, to compare their evolutional behavior and trace their genetic relationship if possible. Indeed, the orbital similarity criterion, $D_{\rm SH}$ \citep{SW} between 2011~$\mathrm{XA_3}$ and Phaethon is 0.196, suggesting that 2011~$\mathrm{XA_3}$ has a possible association range for the orbit of Geminids, because $D_{\rm SH}$ of less than 0.2 indicates the typical associating range. Here we performed the backward and forward numerical integration of the post-Newtonian equation of motion over $\pm 30,000$~yr from initial epoch, applying the ``SOLEX", Ver. 11.01 package, developed by \cite{vitagliano} based on the Bulirsh-Stoer (BS) integrator. 
Then we also computed other possible motions of each asteroid, generating multiple ``clones" at the initial epoch, and integrating them. 
As well as the nominal osculating orbit, other clones had slightly different orbital elements, with the $\pm 1 \sigma$ error based on observational uncertainties. 
We generated the clones on the basis of three possible values (nominal and $\pm 1\sigma$) for five orbital elements [$a$; $e$; $i$; $\mathit{\Omega}$; $\omega$], thus yielding a total of 243 $(=3^5)$ clones, including the nominal one.  Coordinates and velocities of the planets, Moon, and four quasi-planets, Ceres, Pallas, Vesta, and Pluto, regarded as point masses, were based on JPL's DE/LE406-based ephemerides. We confirmed that the results of our numerical integrations did not significantly change when we used other integration methods that we have often applied in our studies, e.g., the Adams method. 
Our integrator can accurately process very close encounters by means of a routine that makes automatic time-step adjustments, and truncation and round-off errors are almost negligible 
for our investigation here. Although the error of the orbital energy in the computation, using the BS method, shows a linear increase with time \citep{chambers}, 
it has an insignificant effect on our integration of $\pm 30,000$~yr. Hence, the SOLEX is rather reliable for dealing with our issue. 
As initial parameters, up to date osculating orbital elements were taken from Nakano's data (personal communication) for 2011~$\mathrm{XA_3}$ and 
the JPL Small-Body Database for Phaethon, as listed in Table~\ref{tbl:XPorbit}. Figure 4 shows the $\pm 30,000$~yr orbital evolutions of 2011~$\mathrm{XA_3}$ and Phaethon. 
Over $60,000$~yr, we found both the asteroids to behave with a high degree of stability with long-period secular changes according to the $\omega$-cycle, i.e., Kozai circulation \citep{Kozai}. 
The corresponding large-amplitude oscillations in $q$-$i$ and antiphase with $e$ occur, thus the period of their cycles  $P_{q} = P_{i} = P_{e}$, 
which is half that of the $\omega$-cycle, $P_{\omega}$ \citep{kino99}, suggesting a typical example of the Kozai circulation.
It is interesting that a spread of the 2011~$\mathrm{XA_3}$ clones looks rather more compact than expected, in spite of preliminary orbital solution of 2011~$\mathrm{XA_3}$, probably due to all the clones being in the stable region of the Kozai mechanism. 
The $\omega$-cycle of $\sim 29,000$~yr for 2011~$\mathrm{XA_3}$ is shorter than that of $\sim 37,000$~yr for Phaethon. 
The time-lag of the orbital evolutions or $\omega$-cycles between 2011~$\mathrm{XA_3}$--Phaethon seems to be near 0 yr, since $\omega$ values of both objects are almost consistent, therefore not so clear as an example of 2005 UD--Phaethon \citep{Ohtsuka06}. 
$D_{\rm SH}$ between 2011~$\mathrm{XA_3}$--Phaethon is presently at around the minimum, in the past $30,000$~yr, however somewhat large when we consider their genetic relation. Also semi major axis of 2011~$\mathrm{XA_3}$ is larger than those of any other PGCs. In addition, our S-complex or V-type taxonomic classification for 2011~$\mathrm{XA_3}$  does not harmonize with the B/F-type taxonomy of Phaethon  \citep{Licandro07}.
Therefore, 2011~$\mathrm{XA_3}$ is not a PGC member. 
\section{Discussion}
\subsection{Relationship with Phaethon}
As we mentioned in the section above, 2011~$\mathrm{XA_3}$ is not a PGC member based on the taxonomic analysis and the orbital calculation. Nevertheless, Phaethon with nominal orbital elements has more opportunities for close encounters with 2011~$\mathrm{XA_3}$ than with either the Earth or the Moon: 59 times within $0.05$ AU in the past $30,000$ years with  2011~$\mathrm{XA_3}$, in comparison with 23 times with the Earth and Moon. This reminds us of the possibility of a collisional event. 
 In recent years, an unexpected brightening and a comet-like dust tail were detected on Phaethon around the perihelion \citep{Li13b, Je13}. This was because thermal fractures and decomposition cracking of hydrated minerals produced the dust ejections, whether the amount of dust production was enough to supply the Geminids in steady state or not was not concluded. The sublimation of ice inside the PGC precursor object due to the thermal evolution has been suggested as a possible mechanism for the breakup of PGC precursor object \citep{Ka09}. Alternatively, impacts event cannot be denied as a possible breakup mechanism. In addition, it is interesting that there exists a rotationally S-type-like color region on Phaethon's surface \citep{cochran84}. Therefore, we infer a $priori$ that  2011~$\mathrm{XA_3}$ might be a remnant candidate impacted with a potential PGC precursor.

\subsection{Surface Material}
Intense solar radiation heating by the perihelion distance of 0.11 AU elevates temperature of the surface material of 2011~$\mathrm{XA_3}$.  Assuming 2011~$\mathrm{XA_3}$ as S-type asteroid $p_{V}$ = 0.209 and $G$ = 0.24, we can estimate the FRM (fast rotating model) surface temperature \citep{Leb89} heating up to 900~K. 2011  $\mathrm{XA_3}$ is heated repeatedly when it comes close to the Sun and the duration at 900~K reaches at least 3,000~yrs (Figure 4), which could result in unique mineralogy of the surface material of 2011~$\mathrm{XA_3}$. A recent sample return mission from S-type asteroid Itokawa revealed that continuous reduction reaction from FeO to metallic Fe due to solar proton implantation to silicates is a main mechanism for developing reduction rims of the silicate crystals on the regolith surface \citep{Noguchi11}. The reduction rims are responsible for space weathering of S-type asteroids. The same reactions are expected to occur on the surface of 2011~$\mathrm{XA_3}$ at a much higher weathering rate, because the asteroid is much closer to the Sun and the reaction takes place at an elevated temperature.
At a low-temperature surface of silicates, the space weathering process makes the reduction rims, consisting of amorphous silicates and metallic iron, from FeO-bearing crystalline silicates \citep{Noguchi11} . On the other hand, at a high-temperature surface of silicates on the asteroid 2011  $\mathrm{XA_3}$, it is expected that the reduction rims thicken, the amorphous silicates crystallize, and the small Fe particles integrate. Our results show that 2011~$\mathrm{XA_3}$ shows a reflectance spectrum feature intermediate between S- and V-type (Figure 2). A meteoritic analog for the surface composition of S-type corresponds to ordinary chondrites that consist mainly of olivine, pyroxene, and plagioclase. The following reaction would take place if olivine were present on the surface:

\begin{equation}
{\rm (Mg,Fe)_2SiO_4} + {\rm H_2} \longrightarrow {\rm MgSiO_3} + {\rm Fe} +{\rm H_2O}.
\end{equation}
Since the reaction proceeds at high temperature, H$_2$O evaporates away from the surface and MgSiO$_3$ crystallizes to pyroxene for a short time. Therefore, pyroxene/olivine ratio and Fe metal abundance are expected to increase on the surface of 2011~$\mathrm{XA_3}$. Interestingly, the increase of pyroxene/olivine ratio is an opposite trend induced by early thermal metamorphism that occurred in the interior of an S-type asteroid 4.6 billion years ago \citep{Gas02}. As for other components of S-type asteroids, pyroxene is more difficult to be reduced than olivine \citep[e.g.,][]{Sin03} and plagioclase does not contain FeO to be reduced. Therefore a major change in mineralogy of the surface of ``S-type" 2011  $\mathrm{XA_3}$ is an increase in the pyroxene/olivine ratio and Fe-metal abundance. If 2011~$\mathrm{XA_3}$ is a V-type asteroid whose meteorite analog is HED (Howardrites, Eucrites, and Diogenites) meteorites, consisting mainly of pyroxene and plagioclase, in this case the mineralogical change is limited to further crystallization of individual minerals. The increase in the pyroxene/olivine ratio during high-temperature space weathering of S-type asteroids makes the mineral assemblage similar to V-type asteroids, which are rich in pyroxene. 2011~$\mathrm{XA_3}$ may be an S-type asteroid with a high pyroxene/olivine or a V-type asteroid. In either case, the reflectance spectrum is similar and difficult to distinguish as we observed in the color-color diagram (Figure 2). 

 Last, we mention the possibility of Q-type because the color-color diagram of 2011~$\mathrm{XA_3}$ indicates the intermediate between S- and V-type, and the population ration of Q-type asteroids is dominant in the NEO region than R- and O-type. 
The cause of 2011~$\mathrm{XA_3}$ being a monolith is thought to the rotational fission of a rubble pile object due to the YORP effect, and the ejector by impact craterings or catastrophic disruption on the parent bodies. As we describe above, the heating to 900~K promotes the space weathering. However, if the rotational fission and the ejection by impacts took place recently, the surface of 2011~$\mathrm{XA_3}$ has not been long exposed to the solar radiation. In that case, the surface color of 2011~$\mathrm{XA_3}$ might indicate Q-type.

\section{Summary}
This study revealed the physical properties of 2011~$\mathrm{XA_3}$ by the photometric observation. We detected the lightcurve periodicity to be 0.0304 $\pm$ 0.0003 day (= 43.8 $\pm$ 0.4 min). The lightcurve amplitude and rotational period clearly deduced 2011~$\mathrm{XA_3}$ to be a monolithic asteroid. The multiband photometric analysis indicated that the taxonomic class of 2011~$\mathrm{XA_3}$ was S-complex, or V-type. Assuming the typical albedo data for S- and V-type, we found the diameter of 2011~$\mathrm{XA_3}$ ranging between 103--323~m, implying the second-largest asteroid among fast-rotating asteroids. We also performed a dynamical simulation for both 2011~$\mathrm{XA_3}$ and Phaethon to be not of common origin. \\

This study ensures the existence of subkilometer-sized fast-rotating monolithic asteroids, of which only a few have been discovered, 2001 OE$_{84}$, 2001 FE$_{90}$ and 2001 VF$_2$. However, the cumulative size-frequency distribution and the other physical properties for subkilometer-sized fast-rotating monolithic asteroids have not been well explained due to the shortage of physical observations of subkilometer-sized asteroids. To detect the fast-rotating monolithic asteroids and deduce the physical properties, the photometric, multiband, and spectroscopic observations should be conducted immediately following the discovery of NEOs, which are listed on the NEO Confirmation Page of Minor Planet Center\footnote{http://www.minorplanetcenter.net/iau/NEO/toconfirm\_tabular.html}. Continuous observations of this kind lead us to clarify the population and size-distribution of subkilometer-sized fast-rotating monolithic asteroids.

\acknowledgments
We acknowledges to S. Nakano for his orbital solution of 2011~$\mathrm{XA_3}$. We also thank N. Takahashi, M. Yoshikawa, and the staff members of Bisei Spaceguard Center for their support for our observation. S. Hasegawa provided us with valuable advice regarding V-type asteroids. We also acknowledge the Japan Space Forum. Detailed and constructive review by Yolande McLean has considerably improved the English presentation of this paper.

\clearpage



\begin{figure}
\epsscale{1}
\plotone{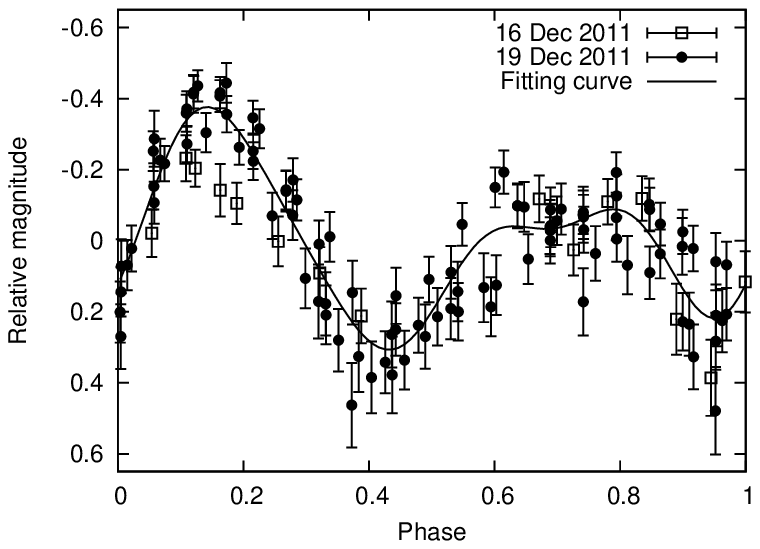}
\caption{Lightcurve of 2011~$\mathrm{XA_3}$. The rotation period is 0.0304 $\pm$ 0.0003 day (= 43.8 $\pm$ 0.4 min). The zero magnitude corresponds to the mean brightness of this asteroid.\label{fig1}}
\end{figure}

\clearpage


\begin{figure}
\begin{center}
\begin{tabular}{c}
{\includegraphics[width=8cm,clip]{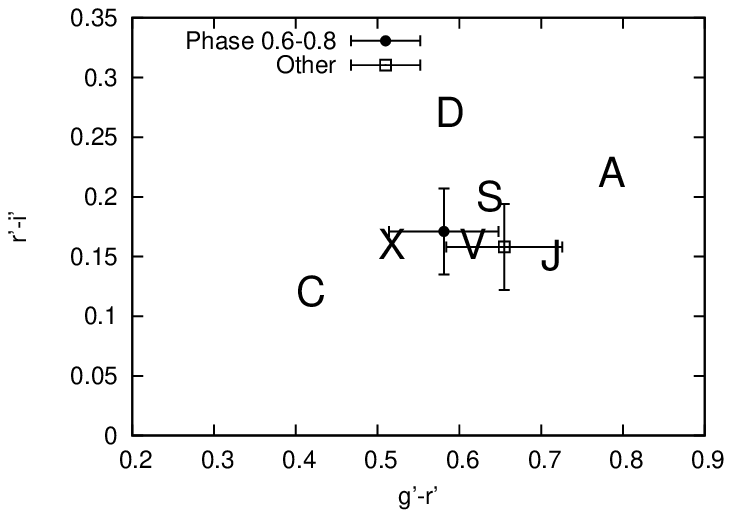}}\\
{\includegraphics[width=8cm,clip]{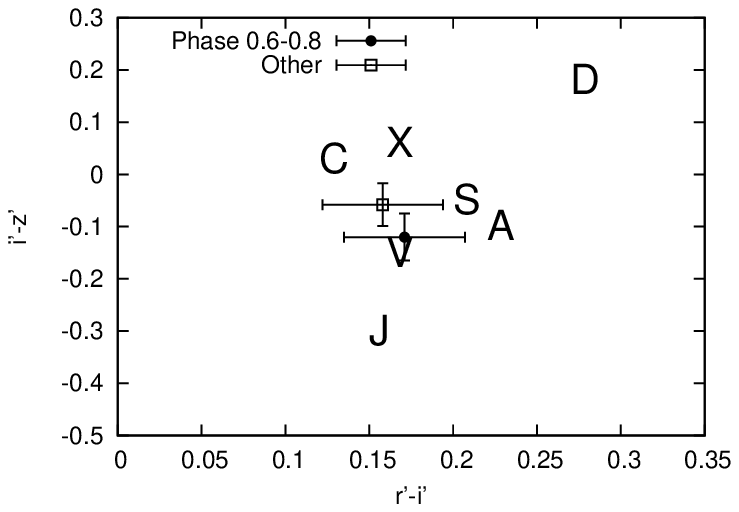}}
\end{tabular}
\end{center}
\caption{Color--color diagram of 2011~$\mathrm{XA_3}$. The large capital letters in the figure represent the taxonomic classes of asteroids on the color--color diagram \citep{Ive01}. The square indicates the averaged color-index between 0.6 and 0.8 in the rotational phase. The filled circle indicates the averaged color-index except for between the phase 0.6 and 0.8.\label{fig2}}
\end{figure}


\begin{figure}
\epsscale{1}
\plotone{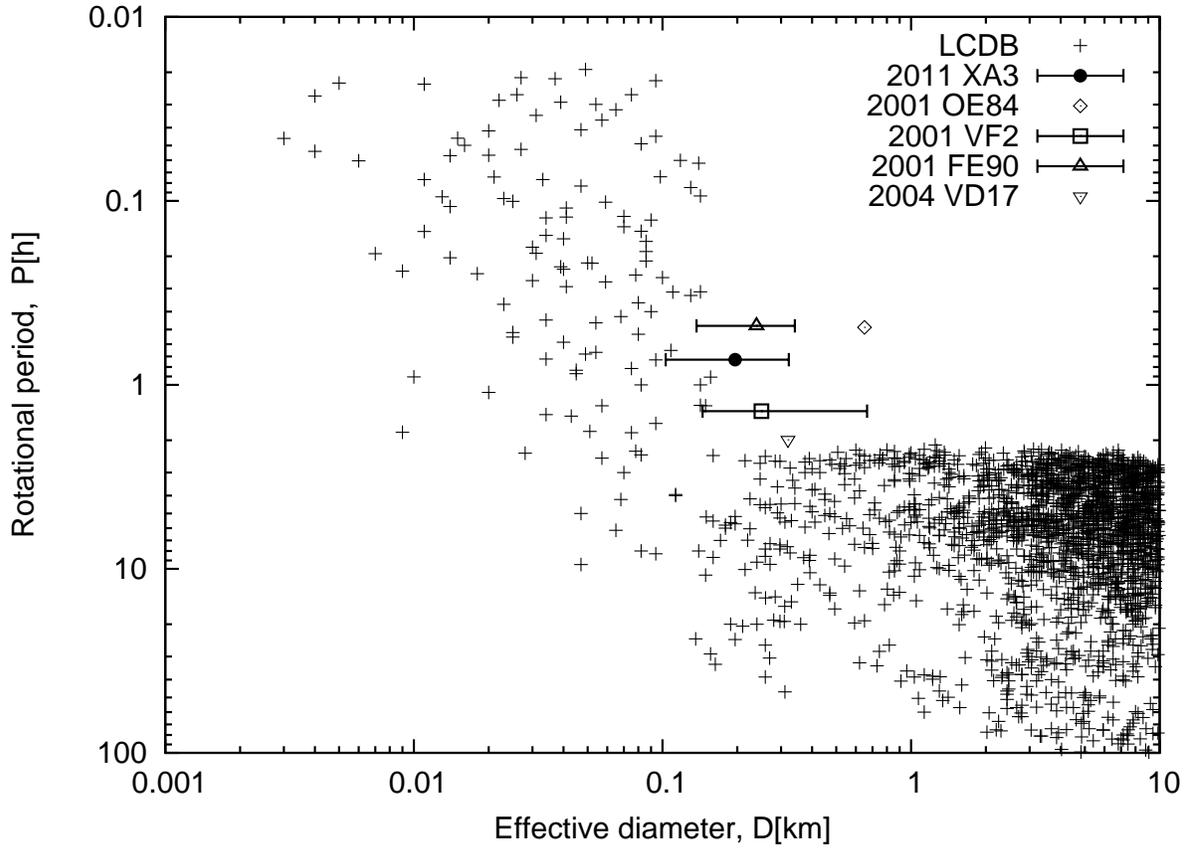}
\caption{Plot of the asteroid rotation period $P$ versus their effective diameter $D$. The legend of `LCDB' shows the data in the Asteroid Lightcurve Data base  with quality code U $\ge$ 2. The filled circle indicates the mean diameter and rotational period of 2011~$\mathrm{XA_3}$. The open marks show the diameter and rotational period for asteroids that have the possibility of being more than 200~m in diameter.}
\end{figure}

\clearpage

\begin{figure}
\begin{center}
\begin{tabular}{c}
{\includegraphics[width=7cm,clip]{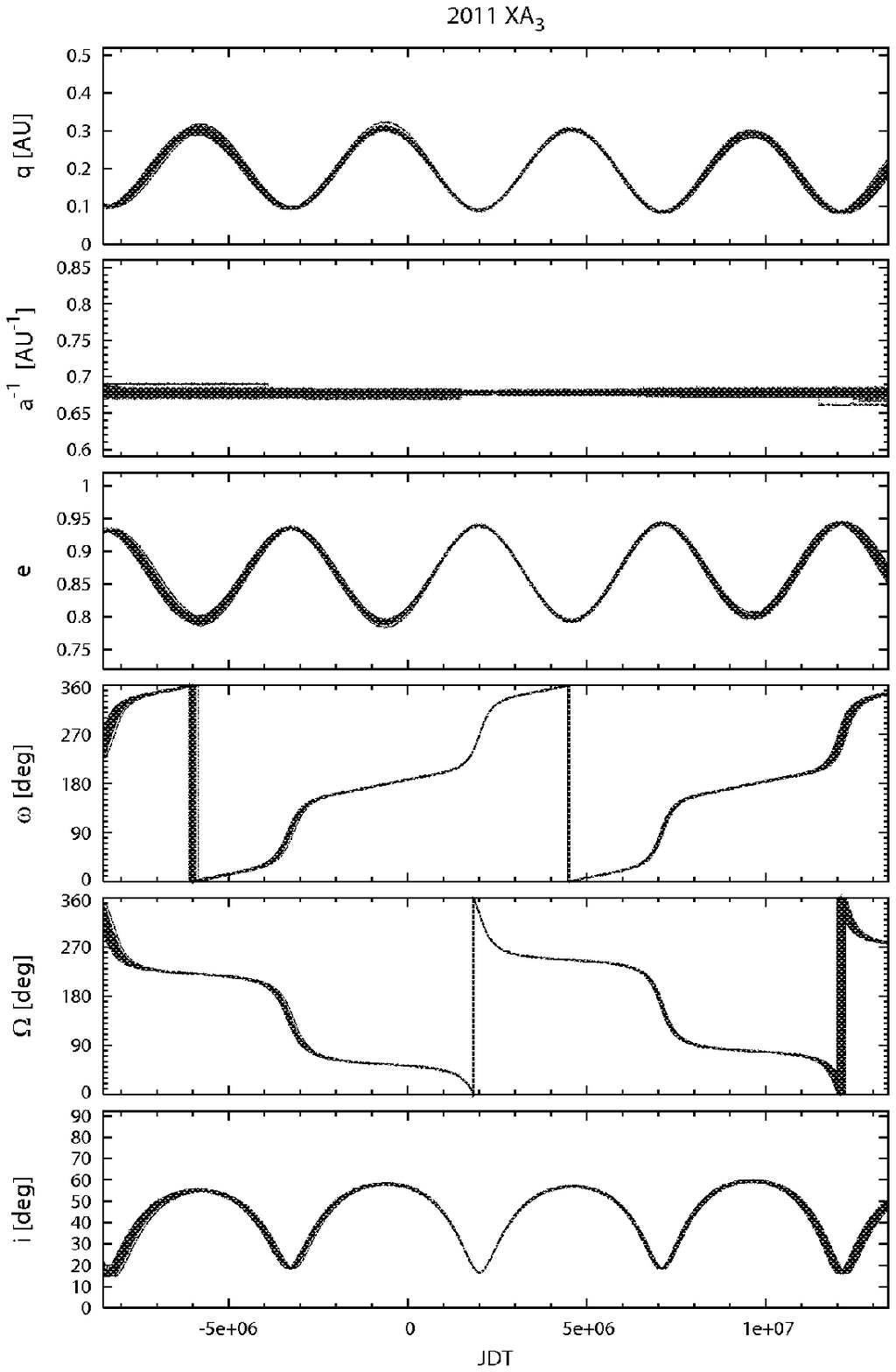}}\\
{\includegraphics[width=7cm,clip]{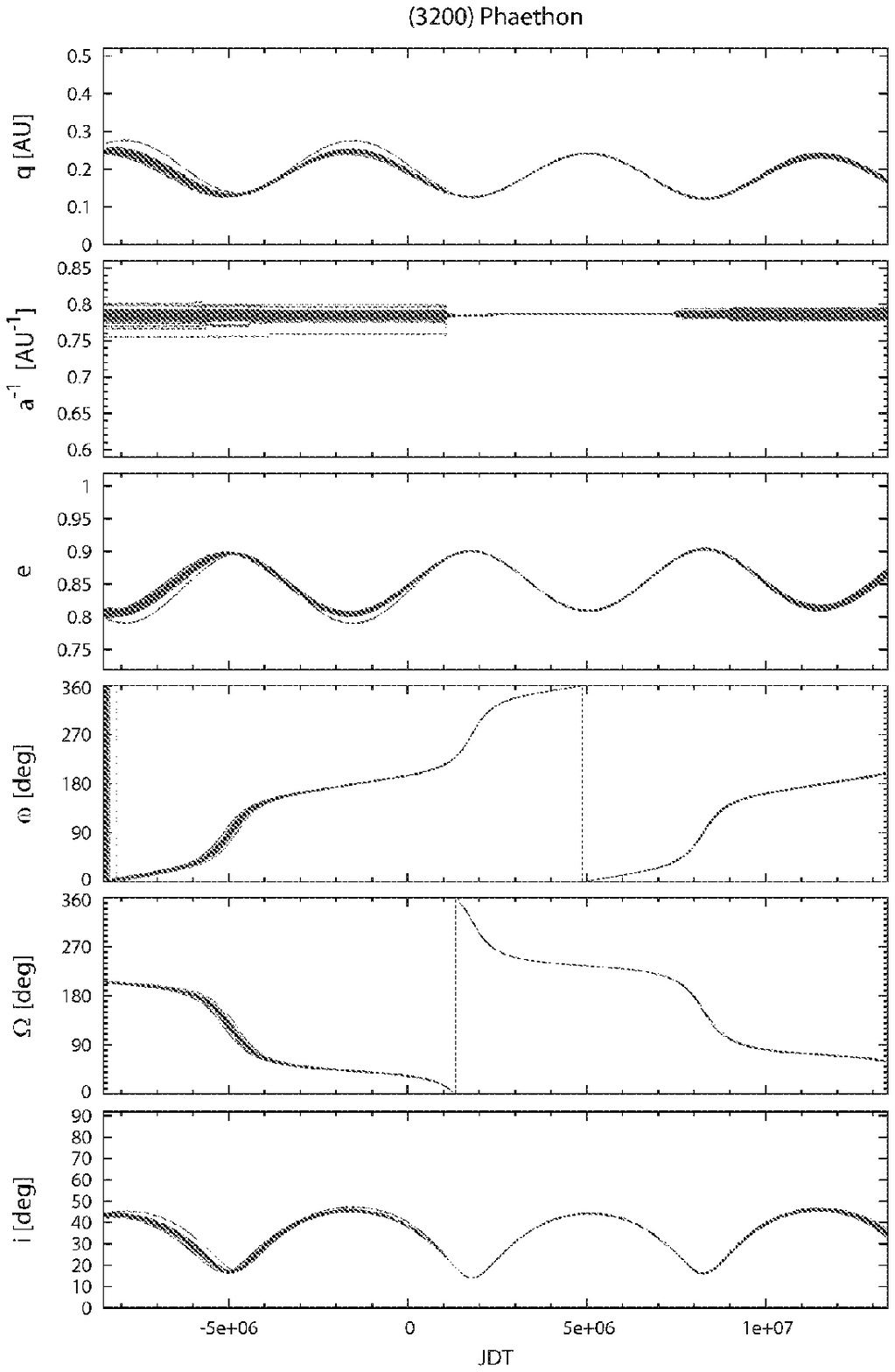}}
\end{tabular}
\end{center}
\caption{Dynamical evolutions for 2011~$\mathrm{XA_3}$ and (3200) Phaethon.\label{fig4}}
\end{figure}







\clearpage

\begin{table}
\caption{Observation states of 2011~$\mathrm{XA_3}$.\label{tbl-1}}
\begin{tabular}{ccccc}
\tableline\tableline
Telescope &  Date [UT]& Exp time(s) &  Filter & Observation term [min]\\
\tableline
BSGC 1.0m & 2011/Dec/16 & 150 &$W$ & $\sim$ 45\\
BSGC 0.5m & 2011/Dec/19 & 120 &$W_{i}$ &$\sim$ 420\\
BSGC 1.0m & 2011/Dec/19 & 120 &$g'$, $r'$, $i'$, $z'$ &$\sim$ 440\\

\tableline
\end{tabular}

\end{table}

\clearpage

\begin{table}
\caption{Geometric configurations of 2011~$\mathrm{XA_3}$. \label{tbl-2}}
\begin{tabular}{ccccc}

\tableline\tableline
Date [UT] &  $R$ [AU]\tablenotemark{a}& $\Delta$ [AU]\tablenotemark{b}  & $\alpha$ [$^{\circ}$]\tablenotemark{c} & Sky motion [$^{\prime\prime}$/min] \\

\tableline
2011/Dec/16.5 & 1.096& 0.141&34.9&6.42\\
2011/Dec/19.5 & 1.044& 0.081&40.8&18.89\\

\tableline
\end{tabular}
\tablenotetext{a}{Heliocentric distance}
\tablenotetext{b}{Geocentric distance.}
\tablenotetext{c}{Phase angle (Sun--2011~$\mathrm{XA_3}$--observer)}

\end{table}

\clearpage

\begin{table}
\caption{Physical properties of 2011~$\mathrm{XA_3}$.\label{tbl-3}}
\begin{tabular}{cc}
\tableline\tableline
Rotational period &  0.0304 $\pm$ 0.0003 day\\
\hline
$g'$ - $r'$ &  0.581  $\pm$ 0.067\tablenotemark{a}\\
$r'$ - $i'$ &  0.171 $\pm$ 0.036\tablenotemark{a}\\
$i'$ - $z'$ &  -0.120 $\pm$ 0.045\tablenotemark{a} \\
\hline
$g'$ - $r'$ &  0.639  $\pm$ 0.070\tablenotemark{b} \\
$r'$ - $i'$ &  0.164 $\pm$ 0.088\tablenotemark{b} \\
$i'$ - $z'$ &  -0.079 $\pm$ 0.054\tablenotemark{b} \\
\hline
 Taxonomy & S-complex $\mid$ V-type \\
 \hline
 Absolute magnitude & 20.56 $\pm$ 0.43  $\mid$ 20.84 $\pm$ 0.25\\
 \hline
 Diameter & 225 $\pm$ 97 m $\mid$ 166 $\pm$ 63 m \\
\hline
Axial ratio & $\ge$ 1.3 $\mid$ $\ge$ 1.4\\
\tableline
\end{tabular}
\tablenotetext{a}{Color index from phase 0.6 to 0.8.}
\tablenotetext{b}{Averaged color index except for between phase 0.6 and 0.8.}
\end{table}

\clearpage

\begin{table*}[htb]
\caption{Initial orbital parameters of 2011~$\mathrm{XA_3}$ and Phaethon including their $\pm 1 \sigma$ error estimates (equinox J2000)}
\label{tbl:XPorbit}
\centering
\begin{tabular}{lcc}
\hline
\hline
Object & 2011~$\mathrm{XA_3}$ & (3200) Phaethon \\
\hline
Osculation epoch (TT) & 2012 Mar 14.0 & 2012 Sep 30.0  \\
Mean anomaly $M$      &  $25\fdg819 \pm 0\fdg074$ & $103\fdg62501443 \pm 0\fdg00000026$ \\
Perihelion distance $q$ (AU) & $0.10746 \pm 0.00012$ & $0.139699845 \pm 0.000000031$ \\
Semimajor axis $a$ (AU) & $1.4753 \pm 0.0028$ & $1.2711609786 \pm 0.0000000021$ \\
Eccentricity $e$ & $0.92716 \pm  0.00019$  & $0.890100587 \pm 0.000000025$ \\
Argument of perihelion $\omega$ & $323\fdg7932 \pm 0\fdg0095$ & $322\fdg1318749 \pm 0\fdg0000097$ \\
Longitude of ascending node $\mathit{\Omega}$ & $273\fdg6070 \pm 0\fdg0033$ & $265\fdg280951 \pm 0\fdg000010$ \\
Inclination $i$  & $28\fdg051\pm 0\fdg029$ & $22\fdg2342789\pm 0\fdg0000076$ \\
Nnumber of astrometric positions &  139 & 2368 \\
Astrometric arc  & 2011 Dec 15-23 (8 days) & 1983--2012 (10364 days) \\ 
RMS residual & $0\farcs28$ &  $0\farcs49$ \\
Reference & personal comm. by S. Nakano & JPL316 \\
\hline
\end{tabular}
\end{table*}

\clearpage







\end{document}